\documentclass[11pt]{article}
\usepackage[preprint]{acl}

\usepackage{times}
\usepackage{latexsym}
\usepackage[T1]{fontenc}
\usepackage[utf8]{inputenc}
\usepackage{microtype}
\usepackage{inconsolata}
\usepackage{graphicx}
\usepackage{amsmath}
\usepackage{booktabs}
\usepackage{multirow}
\usepackage{flushend}

\title{AutoSwarm: Automating Agent Swarms}
\newcommand{\AuthorNames}{Zhonghang Li, Zongwei Li, Chao Huang}
\author{
  Zhonghang Li \quad Zongwei Li \quad Chao Huang \\
  The University of Hong Kong \\
  \texttt{\{bjdwh.zzh, zongwei9888, chaohuang75\}@gmail.com}
}

\hypersetup{pdftitle={AutoSwarm: Automating Agent Swarms},pdfauthor={\AuthorNames}}

\begin{document}
\maketitle
\begin{abstract}
Large language model (LLM) agents increasingly automate complex tasks, with multi-agent systems seeking further performance gains through specialization and collaboration. However, distributing work across agents does not automatically yield complementary contributions, and inter-agent dependencies can turn local errors into task failure. To address these challenges, we propose AutoSwarm, a multi-agent framework for collaborative problem solving. Collaboration formation jointly defines responsibility boundaries and conditions for acceptance and downstream use, grounding capability allocation in contributions that other agents can assess and integrate toward a shared objective. Execution coordination couples agents' local progress through a shared state, using dependencies and supervisory relationships to govern agent activation, result handoffs, and integration while adapting subsequent actions to review feedback and emerging blockers.
Experiments spanning long-horizon programming, interactive planning, and professional financial analysis demonstrate AutoSwarm's advantages over single-agent baselines and existing multi-agent frameworks.
Source code is available at \url{https://github.com/HKUDS/AutoSwarm}.
\end{abstract}

\section{Introduction}

Recent advances in large language models (LLMs) have enabled increasingly capable instruction following and interactive problem solving \citep{ouyang2022instructions,yao2023react}. By combining language-based reasoning with tools and environmental feedback, LLM agents can gather information, formulate plans, and take actions toward user-specified goals. This paradigm supports the automation of tasks that extend beyond a single response, including information seeking, data analysis, and software development. Such tasks require investigation, execution, and verification to work together over a sequence of decisions and actions.

\begin{figure}[t]
\centering
\includegraphics[width=\columnwidth]{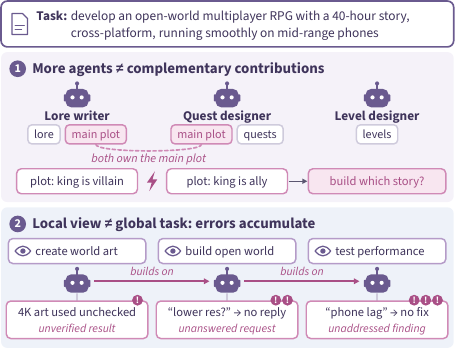}
\caption{Motivation for AutoSwarm: Responsibility Misalignment and Fragile Dependencies}
\label{fig:motivation}
\end{figure}
Many existing LLM agents approach these tasks through a single reasoning-and-action loop. ReAct interleaves reasoning with actions and observations to update plans during interaction \citep{yao2023react}, while SWE-agent equips an agent with interfaces for navigating repositories, editing files, and executing tests \citep{yang2024sweagent}. Such designs enable one agent to handle multiple stages of a task while retaining a unified execution history. However, the same agent remains responsible for gathering evidence, interpreting requirements, and carrying out different kinds of work. As task scope expands, this concentration of responsibilities can limit how thoroughly different parts of the task are addressed. Execution-trace analyses report cases in which single-agent investigation lacks breadth, whereas coordinated agents examine complementary evidence and achieve better performance \citep{kim2025scaling}. These findings motivate using multiple agents to distribute responsibilities and provide complementary contributions.

Multi-agent methods pursue this goal through role specialization, structured workflows, and adaptive orchestration. MetaGPT assigns domain responsibilities and uses standardized procedures to organize work and verify intermediate results \citep{hong2024metagpt}. AgentVerse combines expert recruitment, collaborative decision-making, execution, and evaluation, adjusting its group composition based on feedback \citep{chen2024agentverse}. Magentic-One uses an orchestrator to delegate work to specialized agents, monitor progress, and revise plans when execution encounters difficulties \citep{fourney2024magenticone}. In practice, collaboration remains vulnerable to mismatches between task requirements and work assignments, as well as failures to coordinate dependent activities, which can limit the gains from distributing work across agents \citep{kim2025scaling,cemri2025failures}.

These limitations give rise to two connected challenges, illustrated in Figure~\ref{fig:motivation}. \textbf{First, increasing the number of agents does not automatically yield complementary contributions.} Additional agents create opportunities for specialization and complementary work, but these benefits depend on how responsibilities are assigned and connected. Splitting work by role or step can still leave overlapping responsibilities, missing requirements, or unclear handoffs. The system must define collaboration boundaries according to the task, matching suitable capabilities to concrete responsibilities and specifying the expected outputs and conditions for their acceptance and downstream use. \textbf{Second, dependencies among agents can turn local errors and blockers into task failure.} Once work is distributed, agents hold partial information and progress at different rates, yet rely on one another's results to continue. An unverified result may be adopted downstream, a review finding may never lead to a revision, or an unanswered request may stall dependent work. Coordination must identify such problems as they arise and determine which work can proceed, which needs correction, and who should resolve the blocker.

To address these challenges, we propose AutoSwarm, a multi-agent framework for collaborative problem solving. The framework comprises two components: collaboration formation and execution coordination. Collaboration formation grounds specialization in the contributions needed by the shared objective, defining responsibility boundaries together with the conditions for acceptance and downstream use so that agents' outputs can support one another. Execution coordination maintains a shared state of contribution progress and dependencies to determine the next actions across the team. Independent contributions can proceed concurrently, and accepted results enable dependent work to advance. Review feedback and emerging blockers update the same state, adjusting subsequent actions and identifying the agents responsible for execution, revision, or integration.
We conduct experiments on three benchmarks spanning multiple domains and task types: LongCLI-Bench~\citep{feng2026longcli}, Plancraft~\citep{dagan2025plancraft}, and APEX-Agents 1.1~\citep{vidgen2026apex,bennett2026apex11}. The results show that AutoSwarm improves task completion over single-agent baselines and existing multi-agent frameworks, with consistent gains across different LLM backbones. Our contributions are:

\begin{list}{\textbullet}{  \setlength{\leftmargin}{1em}  \setlength{\labelwidth}{0.5em}  \setlength{\labelsep}{0.5em}  \setlength{\itemsep}{2pt}  \setlength{\parsep}{0pt}  \setlength{\topsep}{3pt}  \renewcommand{\makelabel}[1]{#1\hfil}}
\item We formulate collaboration at the level of interdependent contributions, linking capability allocation to conditions for acceptance and downstream use. This makes the intended use of a result part of its delegation, turning specialization into contributions that other agents can assess and integrate toward the shared objective.
\item We develop execution coordination that makes the conditions for result handoffs govern downstream progress. Review findings and execution blockers update the shared state controlling dependency release, revision, and integration, making correction part of the execution logic and preventing unaccepted results from releasing dependent work.
\item Experiments on three benchmarks covering diverse domains and task types validate the effectiveness of AutoSwarm. The results show higher task completion than single-agent baselines and existing multi-agent frameworks, highlighting the value of task-specific organization and coordinated execution for multi-agent collaboration.
\end{list}

\section{Preliminaries}
\label{sec:preliminaries}

We consider an agentic task described by $(d,\mathcal{E})$, where $d$ specifies the objective and requirements in natural language, and $\mathcal{E}$ is the environment provided for execution. The environment includes the initial context, relevant resources, and available tools, and supports interaction throughout task execution. The required output $y$ is a task result, such as an answer, a generated artifact, or a completed change to the environment.

We describe an agent swarm by a set of available LLM agents $\mathcal{A}$ and a collaboration mechanism $\Omega$. Their collective execution induces the task-level input-output relation
\begin{equation}
 y\sim P_{\mathcal{A},\Omega}(\cdot\mid d,\mathcal{E}),
\label{eq:task-mapping}
\end{equation}
where $P_{\mathcal{A},\Omega}$ denotes the distribution of task results under $\Omega$. Our focus is how collaboration organizes available capabilities and coordinates interdependent work to satisfy the task's requirements.

In AutoSwarm, collaboration is described by an organization $\mathcal{O}$, a delegation graph $\mathcal{G}_t$, and a shared execution state $s_t$, where $t$ indexes coordination updates. Organization $\mathcal{O}$ assigns available agents to task-specific responsibilities and supervisory relationships before execution. Each graph node is a task unit binding a scoped goal to an assigned agent, an expected result, acceptance conditions, and a parent objective; edges encode required result handoffs. State $s_t$ records unit lifecycles, outputs, assessments, terminal outcomes, pending actions, and unresolved requests. Their evolution follows
\begin{equation}
 (\mathcal{G}_{t+1},s_{t+1})
 = \Phi_{\mathcal{O}}(\mathcal{G}_t,s_t;e_t),
\label{eq:collaboration-dynamics}
\end{equation}
where $e_t$ denotes a new delegation, a submitted result, review feedback, or a coordination blocker. The transition $\Phi_{\mathcal{O}}$ adds or revises task units and handoff conditions, uses current evidence to release dependencies or require revision, and assigns pending actions in execution, review, revision, and integration. Unresolved information requests or failed prerequisites prompt supervisory coordination within the same transition. The organization remains fixed while the delegation graph and execution state adapt. Integration combines accepted contributions into $y$ and assesses them against the original requirements. The next section specifies how these relationships are formed and how their evolution is coordinated.

\section{Methodology}
\label{sec:method}

\begin{figure*}[t]
\centering
\includegraphics[width=\textwidth]{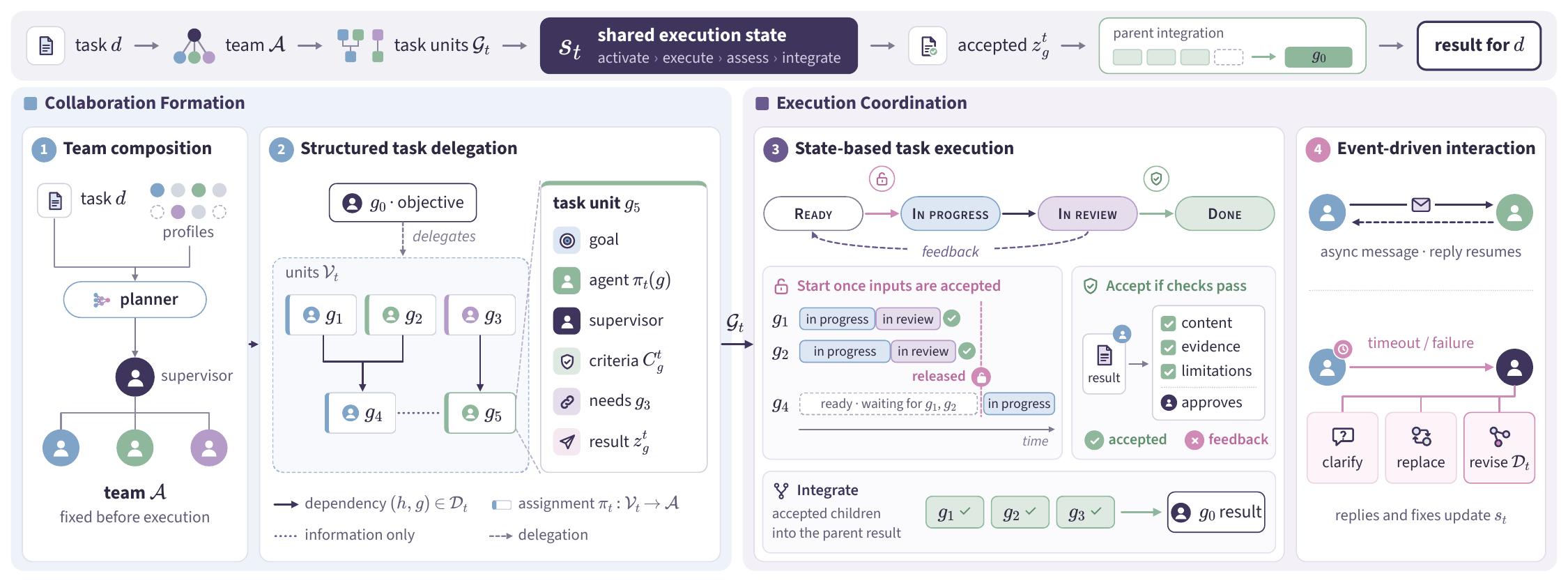}
\caption{Overview of the proposed AutoSwarm framework.}
\label{fig:framework}
\end{figure*}

We propose AutoSwarm, a multi-agent framework comprising collaboration formation and execution coordination, as shown in Figure~\ref{fig:framework}. Collaboration formation connects available capabilities to responsibilities and represents delegated contributions as task units. Execution coordination couples local task execution through dependencies and supervisory relationships. Explicit goal relationships and a shared execution state connect individual reasoning and result handoffs to the overall objective.

\subsection{Collaboration Formation}
\label{sec:collaboration-construction}

Additional agents contribute to a task through distinct, connected responsibilities. AutoSwarm establishes a team around the task's functional needs and translates its responsibilities into task units with scoped goals, explicit outputs, and relationships. The supervisory hierarchy establishes accountability for these goals, while result dependencies specify how contributions support one another.

\subsubsection{Adaptive Team Composition}

A catalogue of agent profiles offers available capabilities, but the task determines which functions are needed and how they should interact. Given the task description $d$ and environment $\mathcal{E}$, a planning model proposes a team according to task complexity, required expertise, and requested responsiveness and output quality. The plan specifies responsibilities, expected contributions, ownership of generated artifacts, required inputs, and supervisory relationships. Each role is justified by the contribution that would be lost if it were removed or merged with another role. This ties team composition to complementary task functions.

Profiles are selected jointly against the proposed responsibilities and their relationships. A suitable profile is retained or refined into instructions for its assigned role, removing irrelevant duties, invented qualifications, incompatible output conventions, and unsupported numerical assumptions. Each role is assigned an agent with the required tools and execution capabilities. Supervisory relationships define the direct-supervisor hierarchy used for delegation and review. The team is established before execution; later adaptation updates delegated goals and their relationships.

\subsubsection{Structured Task Delegation}
\label{sec:structured-delegation}

A role describes expertise, while collaboration requires a scoped contribution that another agent can use. We introduce the \emph{task unit} as the basic unit of delegation: it binds a goal to an assigned agent, an expected result, and the conditions for handing that result over. A supervising agent can create multiple units for an objective, and an agent can receive multiple units associated with different parent objectives. Task assignment therefore relates agents to contributions rather than assigning one agent a single indivisible task.

Each unit has a \emph{responsible agent}, which executes and revises its contribution, and a \emph{supervising agent}, which retains accountability for its parent objective. By default, the supervising agent is the responsible agent's direct supervisor and reviews its submitted result. A specialized reviewer can be explicitly assigned while the supervisor retains integration responsibility. These functions are relative to a unit: an agent can execute the units assigned to it and supervise units delegated to others.

\noindent\textbf{(i) Contribution boundaries.} Each unit specifies its scope, expected output, and ownership of generated artifacts. The assignment identifies the parent requirement it addresses and the inputs it needs. These boundaries assign production and revision to a responsible agent, while the delegating agent remains accountable for integrating contributions.

\noindent\textbf{(ii) Acceptance requirements.} The delegating agent specifies task-specific criteria for required content, supporting evidence, and the assumptions or limitations accompanying the output. These criteria guide execution, review, and correction, connecting the contribution to its downstream use.

\noindent\textbf{(iii) Dependencies and review.} A dependency specifies the result needed for another unit to proceed; a review relationship identifies the agent assessing whether that result meets its acceptance criteria. These relationships separate production, assessment, and consumption while connecting each handoff to explicit criteria for downstream use.

The delegation graph $\mathcal{G}_t=(\mathcal{V}_t,\mathcal{D}_t)$ contains task units as nodes. Assignment is represented by a many-to-one mapping $\pi_t:\mathcal{V}_t\to\mathcal{A}$; $\pi_t(g)$ is the agent responsible for unit $g$. Each unit also records its supervisor, parent objective, output requirements, and acceptance conditions $C_g^t$. An edge $(h,g)\in\mathcal{D}_t$ specifies that executing $g$ requires the accepted result of $h$. Information exchange can occur without creating an execution dependency.

Delegation extends or revises this graph as agents develop their contributions. New units preserve their connection to the parent objective and specify the corresponding outputs and dependencies. A handoff conveys the result, supporting evidence, remaining limitations, and intended use. The supervisor assesses whether the contributions collectively address the objective and can be integrated into a coherent result.

\subsection{Execution Coordination}
\label{sec:execution-coordination}

Distributed responsibilities require productive local execution and consistent progress across connected task units. AutoSwarm coordinates both through the shared execution state: it records what each unit has achieved, what remains unresolved, and which action should follow. The same mechanism connects execution to review and integration, and accommodates blockers arising during interaction.

\subsubsection{State-Based Task Execution}
\label{sec:task-execution}

Tracking progress by task unit connects an agent's local work to the contributions needed elsewhere. Local execution describes one unit's lifecycle; global coordination governs interactions between agents across their units.

\paragraph{Local execution.}
Each unit $g$ follows a lifecycle described by $q_t(g)$:

\noindent\textbf{(i)} \textsc{Ready}. The assigned unit awaits activation for execution or revision.

\noindent\textbf{(ii)} \textsc{In progress}. The responsible agent interprets the goal and uses tools in $\mathcal{E}$ to develop or revise its result. Execution can remain paused while necessary information or inputs are pending.

\noindent\textbf{(iii)} \textsc{In review}. A submitted candidate awaits an assessment decision. Corrective feedback returns the unit to \textsc{Ready} with its prior output and revision instructions.

\noindent\textbf{(iv)} \textsc{Done}. The unit has terminated with success, failure, or cancellation, retaining its output and the recorded outcome.

The unit retains its goal, observations, output, and feedback across turns. Global coordination determines activation, assessment, and inter-agent handoffs.

\paragraph{Global coordination.}
Coordination links agents through their delegated task units, review relationships, and required result handoffs. The shared state $s_t$ gives a joint view of progress across $\mathcal{G}_t$ and determines which agent should execute, assess, or integrate a contribution. Each action receives context assembled from the relevant unit's goal, current inputs, prior output, and feedback.

A designated reviewer, normally the direct supervisor, assesses the contribution and its supporting evidence $z_g^t$. Acceptance is
\begin{equation}
 a_t(g)=C_g^t(z_g^t)\land R_t(g),
\label{eq:result-acceptance}
\end{equation}
where $C_g^t$ checks the output and evidence requirements, $R_t(g)$ records review approval, and $a_t(g)\in\{0,1\}$ indicates acceptance. This decision governs successful closure and the result's availability to other agents. Failed or cancelled units retain $a_t(g)=0$; terminal closure alone does not supply an accepted input.

Required result handoffs determine dependent execution:
\begin{equation}
 \eta_t(g)=\bigwedge_{(h,g)\in\mathcal{D}_t}a_t(h).
\label{eq:dependency-admissibility}
\end{equation}
Global coordination releases a unit from \textsc{Ready} to \textsc{In progress} through this gate, with $\eta_t(g)=1$ and the necessary information available. For task units $g\in\mathcal{V}_t$, result dependencies define the frontier
\begin{equation}
 \mathcal{F}_t=\{g\mid q_t(g)=\text{\textsc{Ready}},\,\eta_t(g)=1\}.
\label{eq:execution-frontier}
\end{equation}
Units without required predecessors satisfy the result condition. Independent eligible units can advance concurrently, and an agent can execute one unit while others assigned to it await input or review. Supervisors can assess child contributions while their parent units remain pending.

Acceptance updates eligibility for connected dependents; new delegations and revised requirements update the graph and shared state together. Revised requirements invalidate prior acceptance and preserve the existing output for reassessment. Integration develops a parent result from accepted child contributions, reconciles their conclusions, and checks parent requirements before submitting it for review. Execution, assessment, revision, and integration thus connect local lifecycles through explicit interactions between agents.

\subsubsection{Event-Driven Interaction}
\label{sec:event-coordination}

Execution reveals information gaps and decisions beyond the declared result dependencies. AutoSwarm addresses these emerging needs through scoped messages and execution events linked to an affected unit, relevant evidence, and the role responsible for the next action. Messages carry their task context and interaction intent, with replies linked to the corresponding requests. Ordinary communication is asynchronous; explicit response requirements associate necessary waiting with the affected action. Actionable messages and execution events activate the relevant agent with the context needed to respond, including roles whose earlier assignments have ended.

Replies and results update the shared state and resume pending actions. Timeouts and unsuccessful contributions activate the supervisor with the unresolved issue and its context. The supervisor coordinates clarification, corrective or replacement contributions, and justified dependency revisions, updating the available inputs, next action, and responsible role together. Failed inputs remain unaccepted, and unrecoverable branches terminate as failures. Appendix~\ref{app:method-details} specifies the communication and supervisory mechanisms.

\section{Experiments}
\label{sec:experiments}

We evaluate AutoSwarm by addressing the following five research questions:
\begin{list}{\textbullet}{  \setlength{\leftmargin}{1em}  \setlength{\labelwidth}{0.5em}  \setlength{\labelsep}{0.5em}  \setlength{\itemsep}{2pt}  \setlength{\parsep}{0pt}  \setlength{\topsep}{3pt}  \renewcommand{\makelabel}[1]{#1\hfil}}
\item \textbf{RQ1:} How does AutoSwarm compare with single-agent systems on long-horizon tasks?
\item \textbf{RQ2:} How does AutoSwarm perform compared with existing multi-agent frameworks?
\item \textbf{RQ3:} How well does AutoSwarm generalize to a specialized professional domain?
\item \textbf{RQ4:} How do collaboration architectures and key mechanisms affect performance?
\item \textbf{RQ5:} How does AutoSwarm improve multi-agent collaboration during execution?
\end{list}

RQ5 case studies appear in Appendix~\ref{sec:rq5}.

\subsection{Experimental Setup}
\label{sec:experimental-setup}

\subsubsection{Benchmarks}
\label{sec:benchmarks}
We use three benchmarks spanning long-horizon programming, interactive planning, and professional financial analysis.

\noindent\textbf{(i) LongCLI-Bench.}
LongCLI-Bench~\citep{feng2026longcli} evaluates long-horizon programming in command-line environments. Its 20 tasks span four engineering categories: implementation from scratch, feature addition, bug fixing, and refactoring. They cover six domains: system programming, web development, data engineering, machine learning, applications, and DevOps. Agents receive a repository and requirements in an isolated environment; executable tests check their code and environment state. Overall Pass Rate, our primary metric, is the fraction passing both Fail-to-Pass (F2P) and Pass-to-Pass (P2P) suites in full. F2P Pass and P2P Pass give the fractions passing each suite in full, assessing requirements and preservation of existing behavior and protected files.

\noindent\textbf{(ii) Plancraft.}
Plancraft~\citep{dagan2025plancraft} evaluates planning in a deterministic Minecraft crafting simulator. Given an initial inventory and target item, agents move and smelt items and observe the updated inventory. Crafting chains vary in complexity; missing resources can make goals infeasible. We evaluate the small test subset of 117 problems: 97 solvable and 20 unsolvable. Task success rate is the fraction producing the target item or correctly identifying an infeasible goal. Overall success rate covers all problems; Solvable and Unsolvable success rates cover their respective subsets.

\noindent\textbf{(iii) APEX-Agents 1.1.}
We select the \emph{Investment Banking Analyst} subset of APEX-Agents 1.1~\citep{vidgen2026apex,bennett2026apex11}. Its 80 tasks require finding financial inputs and analytical conventions across spreadsheets, documents, and communications, then computing results or creating and updating financial artifacts. Tasks provide prompts, files, and application tools in a project environment. A judge evaluates outputs and artifacts against expert-authored binary rubric criteria. Mean Score averages the proportion of satisfied criteria across tasks; Pass@1 is the fraction satisfying all criteria in one attempt.

\subsubsection{Baselines and Configurations}
\label{sec:baselines}

\noindent\textbf{(i) Single-agent comparisons.}
For RQ1, we compare AutoSwarm with Claude Code using MiniMax-M3 and DeepSeek-V4-Flash, and with Codex using GPT-5.6-Luna at max reasoning effort. AutoSwarm organizes multiple instances of the corresponding baseline agent with the same backbone. Each configuration is evaluated twice on all 20 tasks, and we report arithmetic mean pass rates. We also include reference results from \citet{feng2026longcli}: GPT-5.3-Codex with Codex, Claude-Opus-4.6 with Claude Code, and DeepSeek-V3.1, GLM-4.6, and Qwen3-235B-A22B with OpenHands.

\noindent\textbf{(ii) Multi-agent comparisons.}
For RQ2, we compare AutoSwarm with AutoGen~\citep{wu2024autogen}, MetaGPT~\citep{hong2024metagpt}, Teamwork, the team-discussion framework of \citet{pappu2026teams}, and Codex (ultra), which uses Codex's built-in subagent orchestration~\citep{openai2026models}. Codex at max provides a single-agent reference. All RQ2 configurations use GPT-5.6-Luna at max reasoning effort, except Codex (ultra), which uses ultra mode; each is evaluated once on all 117 problems.

\noindent\textbf{(iii) Professional-domain references.}
For RQ3, AutoSwarm uses GPT-5.6-Luna with max reasoning effort and is run once on each of the 80 tasks. We compare it with official reference scores for GPT-5.6-Luna, DeepSeek-V4.1-Flash, Gemini 3.1 Pro, GLM-5.1, GPT-6-Luna, MiniMax-M3, and Muse Spark 1.2 on the same subset~\citep{mercor2026banking}.

\noindent\textbf{(iv) Collaboration architectures.}
For RQ4, we implement four architectures adapted from \citet{kim2025scaling}. Independent agents operate in parallel in a shared environment without member message exchange or an LLM aggregator. Decentralized uses three rounds of all-member discussion, followed by voting over final answer strings and submission by an existing member. Centralized uses one coordinator and three members with star-shaped communication. Hybrid adds one peer exchange after the initial member execution to the centralized procedure. All configurations use GPT-5.6-Luna with max reasoning effort and are evaluated once on the 117 Plancraft-small problems.

\subsection{Long-Horizon Performance (RQ1)}
\label{sec:rq1}

\begin{figure*}[t]
\centering
\includegraphics[width=\textwidth]{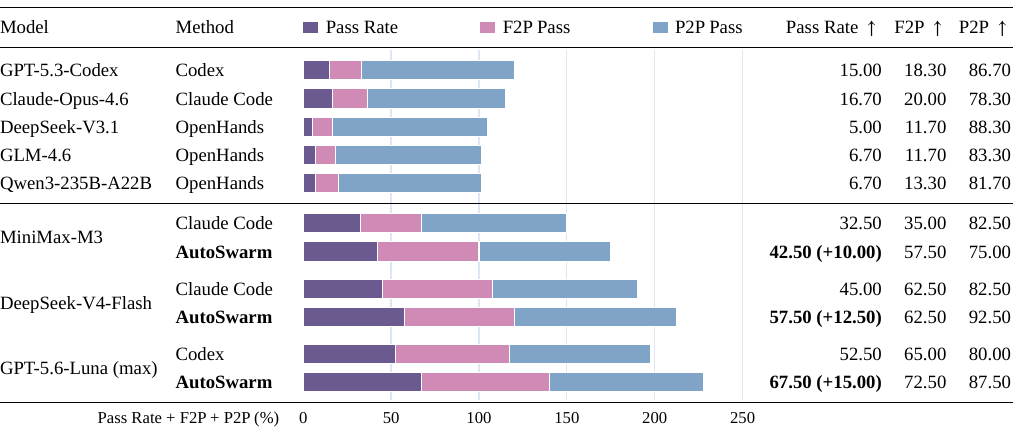}
\caption{LongCLI-Bench results (\%).}
\label{fig:longcli-results}
\end{figure*}

To assess whether AutoSwarm improves completion over extended execution, we compare it with baseline agents on LongCLI-Bench. Figure~\ref{fig:longcli-results} reports two-run means and published references.

\paragraph{Long-horizon gains.}
AutoSwarm improves overall Pass Rate by 10.0, 12.5, and 15.0 percentage points for MiniMax-M3, DeepSeek-V4-Flash, and GPT-5.6-Luna, respectively, averaging 12.5 points. Gains recur in both runs for each backbone. These results support its combined collaboration formation and execution coordination: contribution boundaries and acceptance requirements align delegated work with the objective, while shared-state review, revision, and dependent handoffs connect contributions to joint F2P/P2P completion.

\paragraph{Adaptability across models.}
Overall gains across three backbones support the adaptability of AutoSwarm's responsibility assignment and shared-state execution control. DeepSeek-V4-Flash improves P2P Pass by 10.0 points while maintaining F2P Pass, and GPT-5.6-Luna improves both by 7.5 points, suggesting that review and dependency-aware integration help reconcile implementation and preservation requirements. MiniMax-M3 achieves a 22.5-point F2P gain with a slight P2P decline. Trace analysis attributes this to more frequent protected-file modifications; newly completed tasks more than offset the losses.

\subsection{Multi-Agent Comparison (RQ2)}
\label{sec:rq2}

\begin{figure}[t]
\centering
\includegraphics[width=\columnwidth]{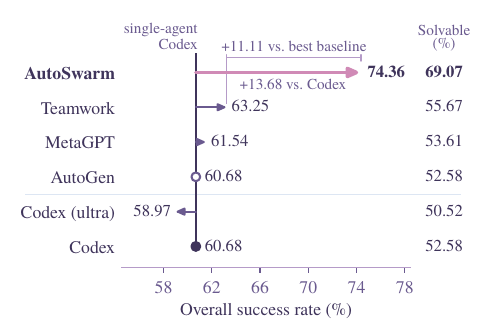}
\caption{Plancraft-small success rates (\%).}
\label{fig:plancraft-results}
\end{figure}

To assess gains beyond existing agent systems, we compare AutoSwarm with Codex configurations and multi-agent methods on Plancraft-small (Figure~\ref{fig:plancraft-results}). AutoSwarm reaches 74.36\% success, improving on Codex (60.68\%) by 13.68 percentage points, while Codex (ultra) falls to 58.97\%. Ultra's decline highlights the challenge of aligning distributed contributions with shared context and dependencies. AutoSwarm governs execution through accepted results and dependency status.

Compared with existing multi-agent methods, AutoSwarm outperforms AutoGen, MetaGPT, and Teamwork, exceeding the strongest baseline, Teamwork, by 11.11 points. MetaGPT emphasizes SOP-based role workflows~\citep{hong2024metagpt}, while Teamwork uses discussion and answer aggregation~\citep{pappu2026teams}. AutoSwarm couples delegation and acceptance criteria with dynamically updated dependencies and assessed result handoffs. All configurations correctly identify the 20 unsolvable problems, so gains come entirely from solvable goals: 69.07\% success for AutoSwarm versus 55.67\% for Teamwork, supporting coordination that carries contributions through execution.

\subsection{Professional Generalization (RQ3)}
\label{sec:rq3}

\begin{table}[t]
\centering
\caption{APEX-Agents 1.1 banking results (\%).}
\label{tab:apex-results}
\small
\renewcommand{\arraystretch}{1.12}
\setlength{\tabcolsep}{3pt}
\begin{tabular*}{\columnwidth}{@{\extracolsep{\fill}}lcrr@{}}
\toprule
Model & Effort & Mean Score $\uparrow$ & Pass@1 $\uparrow$ \\
\midrule
DeepSeek-V4.1-Flash & Max & 43.20 & 38.80 \\
Gemini 3.1 Pro & High & 45.20 & 38.00 \\
GLM-5.1 & -- & 46.80 & 39.60 \\
GPT-6-Luna & Max & 45.50 & 38.80 \\
MiniMax-M3 & High & 42.50 & 36.90 \\
Muse Spark 1.2 & xHigh & 31.50 & 23.40 \\
\midrule
GPT-5.6-Luna & Max & 40.90 & 35.30 \\
\textbf{AutoSwarm} & Max & \textbf{46.99} & \textbf{41.25} \\
\bottomrule
\end{tabular*}
\end{table}

To assess professional generalization, we evaluate AutoSwarm on the Investment Banking Analyst subset of APEX-Agents 1.1 against official model references (Table~\ref{tab:apex-results}).

AutoSwarm improves Mean Score and Pass@1 over the official GPT-5.6-Luna reference by 6.09 and 5.95 percentage points, respectively. It achieves the highest listed Pass@1 (41.25\%), exceeding GLM-5.1 by 1.65 points. Higher complete rubric fulfillment highlights joint requirement satisfaction: explicit contribution requirements and dependent handoffs connect evidence across artifacts, while integration checks whether accepted outputs satisfy the overall objective. These results support generalization beyond programming and planning to professional analysis.

\subsection{Collaboration Design (RQ4)}
\label{sec:rq4}

\begin{figure}[t]
\centering
\includegraphics[width=\columnwidth]{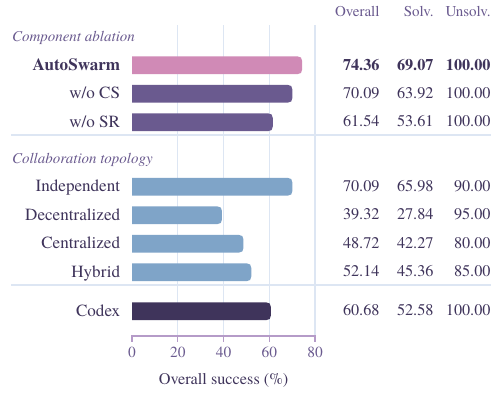}
\caption{Plancraft-small ablation results (\%).}
\label{fig:rq4-results}
\end{figure}

To assess collaboration design, we compare four architectures adapted from \citet{kim2025scaling} and two AutoSwarm variants on Plancraft-small (Figure~\ref{fig:rq4-results}). The architecture comparison examines different communication and control structures, while the mechanism ablations assess how explicit contribution requirements and supervisory review support coordinated execution. The variants remove \emph{contribution specifications} (CS), which define outputs, supporting evidence, and acceptance criteria, or \emph{supervisory review} (SR), which assesses results before integration. Other mechanisms remain; see Appendix~\ref{app:rq4-configurations}.

\paragraph{Architecture comparison.}
Independent is the strongest architectural baseline (70.09\%), outperforming Codex (60.68\%) without member communication. In contrast, Decentralized, Centralized, and Hybrid attain 39.32\%, 48.72\%, and 52.14\%, respectively, all below Codex. Adding peer exchange to Centralized improves success by 3.42 points, but remains insufficient to close the gap to Independent. Under this evaluation protocol, richer communication and centralized control alone do not ensure effective collaboration. AutoSwarm reaches 74.36\%, surpassing Independent by 4.27 points overall and 3.09 points on solvable problems, while correctly identifying all infeasible goals. Rather than relying on communication topology alone, its contribution specifications clarify what agents must deliver, while shared execution state and assessed handoffs govern how their results advance dependent work.

\paragraph{Mechanism ablations.}
Removing CS lowers success to 70.09\%, with five fewer solvable problems completed despite retaining the team, goals, and dependencies. Predefined outputs and evidence criteria help execution and review produce contributions suited to downstream use. Removing SR reduces success by 12.82 points to 61.54\%, near Codex's 60.68\% despite preserved contribution specifications. This loss highlights how assessment triggers revision and governs which results enter integration and dependent work. Both variants retain 100\% unsolvable success, locating their impact in completing feasible goals.

\section{Conclusion}
\label{sec:conclusion}

We presented AutoSwarm, a framework for improving LLM agent performance by turning individual capabilities into coordinated contributions toward a shared objective. Collaboration formation establishes distinct responsibilities and specifies what each agent should produce, together with the conditions for accepting and using its results. Building on these conditions, execution coordination links contribution progress and dependencies through a shared state to guide execution, review, revision, and integration. Feedback and emerging blockers update this state, allowing the team to adjust subsequent actions as execution unfolds. Evaluations across three benchmarks show higher task completion than single-agent baselines and existing multi-agent frameworks, with consistent gains across different LLM backbones.

\bibliography{references}
\flushcolsend
\clearpage
\appendix
\section{Supplementary Material}
\label{app:supplementary}

This appendix provides additional methodological details, reviews related work, and documents supplementary experimental configurations. It concludes with paired case studies illustrating how AutoSwarm's collaboration mechanisms affect execution.

\subsection{Methodological Details}
\label{app:method-details}

This section expands AutoSwarm's interaction mechanisms underlying Sections~\ref{sec:structured-delegation} and~\ref{sec:execution-coordination}. These mechanisms associate communication with task responsibilities and connect emerging coordination needs to the shared execution state.

\subsubsection{Asynchronous Communication}
\label{app:communication}

Agents communicate through persistent, role-addressed messages. A message identifies its sender and recipient, the associated task unit or parent objective, its interaction intent, and its content. It also retains its sending time, response requirement, and any connection to an earlier request. The content provides the question, decision, progress update, or result notification together with the evidence or artifact references needed to interpret it. These associations preserve the link between an exchange and the contribution it concerns when an agent participates in several task units.

Ordinary messages are delivered asynchronously, allowing the sender to continue independent work. A request can explicitly require a response before a particular decision proceeds. This requirement establishes a response obligation and records the pending question, intended recipient, waiting interval, and continuation or supervisory pathway. Replies reference the original request, so arrival is checked against the unresolved exchange. Acknowledging receipt records that a message was seen; satisfying an explicit response obligation requires a linked reply. Both sent and received correspondence remain available for later reasoning, review, and supervisory decisions.

\subsubsection{Role Activation and Context}
\label{app:activation}

Incoming messages and execution events are associated with the role capable of taking the next action. Informational updates contribute to the relevant context, whereas requests, required review, and unresolved decisions create pending response or supervisory actions. These actions retain their unit and parent-objective associations. The context supplied for an action includes the current objective, expected contribution, available results, relevant correspondence, and any prior output or correction feedback. This ties the agent's next reasoning step to the specific obligation it needs to address.

Activation follows the continuing responsibilities of a role. A new actionable exchange can activate an agent after its previous assignment has ended, using the role's retained context and the new request. An occupied agent receives the pending action in its subsequent execution context; other eligible agents can continue their work. The role's activation and the lifecycle of a particular contribution are tracked separately, so a new response obligation can be handled without treating every completed contribution as unfinished. Message associations and recent interaction history identify repeated deliveries or recurring exchanges, providing guards against redundant activation.

\subsubsection{Waiting and Supervisory Decisions}
\label{app:waiting}

Explicit waiting is attached to the information needed for a pending action. The waiting record retains the original request, expected respondent, unresolved response requirements, and the context for continuation. Matched replies are added to the requesting agent's context and resolve the corresponding obligations. Execution remains suspended while required replies are outstanding; unrelated execution, assessment, and integration remain eligible. Before establishing a required wait, the mechanism checks that the recipient can respond without first depending on completion of the requesting contribution. This check addresses circular waiting introduced by an unsuitable recipient relationship.

Configured waiting intervals support timeout handling. Expiration preserves the unresolved question and elapsed waiting context for the configured continuation or supervisory action. A continuation carries the unresolved limitation into subsequent reasoning; it does not mark a missing required result as accepted. Supervisory attention is directed to the responsible agent's superior with the original objective, requested information, intended respondent, and available evidence. The supervisor can seek clarification, assign a replacement contribution, or revise an obsolete dependency while retaining accountability for the parent requirements. Its decision determines the next responsible role and updates the affected unit's pending action and available inputs.

\subsubsection{Review and Dependency Updates}
\label{app:review-dependencies}

Result handoffs connect an output to its parent requirement and acceptance conditions. The review context contains the candidate result, supporting evidence, artifact references, assumptions, and remaining limitations. The reviewer is normally the responsible agent's direct supervisor; an explicitly assigned specialist can assess the result while the supervisor retains integration responsibility. Review feedback identifies deficiencies relative to the required contribution, and revision resumes with the previous output and the requested corrections. Contributions specified to omit separate review retain their output and evidence checks, represented by $R_t(g)=1$ in Equation~\ref{eq:result-acceptance}.

Acceptance releases only the result dependencies that are satisfied under the current graph and unit requirements. Additional required inputs suspend affected execution, and changes to the required contribution invalidate prior acceptance for reassessment. Failed or cancelled units have terminal outcomes but supply no accepted ordinary input. Their partial evidence and failure context support a supervisory decision once the remaining required contributions have completed or cannot progress. The supervisor can arrange replacement contributions, explicitly revise obsolete dependency relationships, or close branches that cannot satisfy the requirements. Ordinary dependent execution remains subject to Equation~\ref{eq:dependency-admissibility}; activating the supervisor for a decision does not itself release these dependencies.

\subsection{Related Work}
\label{sec:related-work}

\subsubsection{LLM Agents}

LLM agents extend language generation into interactive problem solving by combining reasoning, tool use, and environmental observations. Early approaches establish this connection through interleaved reasoning and acting, learned tool invocation, or executable code as a unified action space \citep{yao2023react,schick2023toolformer,wang2024codeact}. Subsequent work strengthens sequential decision-making through feedback-based reflection and search over alternative trajectories \citep{shinn2023reflexion,zhou2024lats}. Experience can also guide later execution: reusable workflow memories abstract common routines from completed interactions \citep{wang2025awm}. In parallel, interface design grounds language-model decisions in concrete environments. Software agents use repository-oriented commands to navigate, edit, and test code \citep{yang2024sweagent}, while multimodal web agents combine visual observations with executable browser actions \citep{zheng2024seeact,he2024webvoyager}. More recent work aligns observation and action representations with the language model's capabilities to improve navigation \citep{yang2025agentoccam}. However, concentrating investigation, implementation, and verification within one agent can still limit coverage as task requirements grow, motivating task-specific collaboration beyond a single execution process.

\subsubsection{Multi-Agent Systems}

Multi-agent systems improve task-solving effectiveness through specialization, interaction, and adaptive orchestration. Role-conditioned dialogue and configurable conversation patterns provide abstractions for cooperative problem solving \citep{li2023camel,wu2024autogen}. Organization-inspired workflows further connect specialized stages through standardized outputs and communication protocols \citep{hong2024metagpt,qian2024chatdev}, while adaptive recruitment aligns expert teams with the current goal and evaluation feedback \citep{chen2024agentverse}. Graph-based approaches make the collaboration structure itself a modeling target: GPTSwarm optimizes prompts and connectivity, G-Designer generates task-aware communication topologies, and MacNet orchestrates interactions over directed acyclic networks \citep{zhuge2024gptswarm,zhang2025gdesigner,qian2025macnet}. Automated design extends this idea to searching executable workflows and sampling query-dependent architectures from an agentic supernet \citep{zhang2025aflow,zhang2025maas}, allowing collaboration structures to be selected according to the problem. Despite advances in team and communication design, collaboration remains vulnerable to misaligned responsibilities and insufficiently verified handoffs \citep{cemri2025failures}, motivating AutoSwarm's explicit connection between contribution requirements and dependency-controlled execution.

\subsection{Additional Experimental Details}
\label{app:experimental-details}
\subsubsection{Collaboration Configurations}
\label{app:rq4-configurations}

\paragraph{Architectural comparisons.}
We adapt the four architectures in \citet{kim2025scaling} to the shared Plancraft environment. Independent agents execute in parallel without member message exchange or an LLM aggregator; their actions affect the common environment state. Decentralized uses three rounds of all-member discussion, followed by voting over final answer strings and submission by an existing member. Centralized uses one coordinator and three members with star-shaped communication. Hybrid follows the centralized procedure and adds one peer exchange after the initial member execution. These settings compare complete collaboration procedures, including their communication and result-submission policies.

\paragraph{Mechanism variants.}
The variant without CS replaces explicit deliverable requirements, supporting-evidence requirements, and acceptance conditions with ordinary goal descriptions. It retains the original team, goal assignments, dependencies, and execution process. Review remains active but proceeds without predefined contribution standards. The variant without SR retains contribution specifications and the other execution mechanisms; supervisors integrate submitted contributions without reviewing them first, eliminating rejection and corrective feedback from this assessment stage. Both variants use the same benchmark instructions and success criteria as the complete framework. Each configuration uses GPT-5.6-Luna with max reasoning effort and is evaluated in one complete run on all 117 problems.

\subsection{Collaboration Case Study (RQ5)}
\label{sec:rq5}

To examine how AutoSwarm's mechanisms operate during execution, we inspect five selected paired trajectories: three Plancraft problems and two LongCLI-Bench tasks. Each pair compares AutoSwarm with a single agent using the same backbone and tools. The Plancraft examples come from the hint-free case-study run with a 30-step environment limit.

\paragraph{Contribution coverage.}
Explicit responsibility boundaries and output requirements make specialization useful to the shared objective. In the piston case, a recipe specialist supplied a complete material-conversion and assembly plan that the supervising agent used to reach the goal; the single agent confused crafting with smelting. In the DeepSeek-V4-Flash buffer-pool case, the single agent implemented the buffer core and page guards but left the required LRU-K replacer as an empty stub, indicating a gap in tracking completion across interrelated modules. AutoSwarm assigned the replacer, page guards, and buffer core to distinct owners, with their outputs independently tested during integration. These handoffs provided both reusable evidence and coverage of required functionality.

\paragraph{Dependency-aware execution.}
Separating information gathering from state modification made resource use follow established prerequisites. In the sticky-piston case, the single agent smelted the only log into charcoal and then incorrectly declared the goal infeasible. AutoSwarm's recipe contribution established the plank prerequisite, and the supervising agent converted the log accordingly before assembly. The benefit arose in execution ordering: consuming prerequisite evidence before acting preserved a scarce input for the feasible plan.

\paragraph{Verification and completion.}
The MiniMax-M3 Scheme case shows how verification supports correction even when initial reasoning is flawed. The assigned specialist initially reproduced the single agent's duplicate-binding error, but focused tests exposed it, prompting a repair that the supervising agent independently confirmed. The golden-chestplate case extends checking to the final environment state: the single agent stopped with the target in the crafting output slot, while AutoSwarm's plan required collection into inventory before completion. Scoped testing, supervisory confirmation, and explicit completion requirements therefore addressed both implementation defects and premature termination.

\end{document}